\documentclass[12pt]{article}
\usepackage{graphicx}
\usepackage{color}
\usepackage{bm,amsfonts,amssymb,epsfig}
\def\be{\begin{equation}}
\def\ee{\end{equation}}
\def\bea{\begin{eqnarray}}
\def\eea{\end{eqnarray}}
\definecolor{blueish}{rgb}{0.0,0.0,0.7}
\definecolor{greenish}{rgb}{0.0,0.7,0.0}
\definecolor{darkgreen}{rgb}{0.0,0.4,0.0}
\definecolor{turqoise}{rgb}{0.0,0.5,0.5}
\definecolor{gold}{rgb}{0.7,0.6,0.0}

\begin{document}
\title{MULTICHANNEL DESCRIPTION OF \\ LIGHT AND INTERMEDIATE SCALAR MESONS}
\author{
Susana Coito\footnote{E-mail: susana.coito@ist.utl.pt} ~and
George Rupp\footnote{Invited speaker; e-mail: george@ist.utl.pt} \\
{\small Centro de F\'{\i}sica das Interac\c{c}\~{o}es Fundamentais,
Instituto Superior T\'{e}cnico}, \\ {\small Technical University of
Lisbon, P-1049-001 Lisboa, Portugal} \\[2mm]
Eef van Beveren\footnote{E-mail: eef@teor.fis.uc.pt} \\ {\small
Centro de F\'{\i}sica Computacional, Departamento de F\'{\i}sica,
Universidade de Coimbra}, \\ {\small P-3004-516 Coimbra, Portugal}
\mbox{ } \\[-8mm] }
\date{}

\maketitle

\begin{abstract}
The light scalar mesons in a variety of approaches are briefly reviewed,
as well as their description in the Resonance-Spectrum Expansion and related
coupled-channel formalisms. A recent multichannel modelling of the light 
scalars is extended to higher energies and with additional decay channels,
allowing to make predictions for the intermediate scalar mesons as well.
Prospects for further improvements are discussed. \\[2mm]
\end{abstract}

\section{Introduction to Scalar Mesons} 
\mbox{ } \\[-7mm]
The light scalar mesons represent nowadays one of the hottest topics in
hadronic physics. Despite the growing consensus on the existence of a
complete light scalar nonet, comprising the $f_0$(600) (alias $\sigma$),
$K_0^*$(800) (alias $\kappa$), $a_0$(980) and $f_0$(980), which are now all
included \cite{PDG08} in the PDG tables, their interpretation and possible
dynamical origin in the context of QCD-inspired methods and models remains
controversial. Moreover, their classification with respect to the intermediate
scalars $f_0$(1370), $K_0^*$(1430), $a_0$(1450) and $f_0$(1500) \cite{PDG08}
is also subject to continued debate. Thus, before presenting our actual model
calculations, a brief historical discussion of the main theoretical and
phenomenological approaches to the light scalars appears quite opportune,
facing this audience that covers many different fields of expertise.
Of course, time and space limitations do not allow an exhaustive treatment
here.

What makes the light scalars so awkward for quark-model builders is not only
their ``lightness'', as their masses would rather be expected in the
1.3--1.5 GeV region, for conventional $^{3\!}P_0$ $q\bar{q}$ states, but also
the fact that the isoscalar $\sigma$ meson is much lighter than the isovector
$a_0$(980). Moreover, the obviously nonstrange quark content of the $a_0$(980)
makes it very difficult to understand its approximate mass degeneracy with the
$f_0$(980), which is dominantly an $s\bar{s}$ state, as can be inferred e.g.\
from the non-observation \cite{PDG08} of the decay $a_1(1260)\to f_0(980)\pi$,
whereas the process $a_1(1260)\to\sigma\pi$ \em is \em \/seen \cite{PDG08}.

These three problems were seemingly solved simultaneously in Jaffe's \cite{J77}
$qq\bar{q}\bar{q}$ proposal for ground-state scalar mesons, more than
30 years ago. Namely, a very strong and attractive colour-hyperfine
interaction for the lowest scalar $q^2\bar{q}^2$ (alias tetraquark) states
could explain their low masses, while the mass degeneracy of the $a_0$(980)
and $f_0$(980) would follow naturally from both having an
$n_i\bar{n_j}s\bar{s}$ configuration, where $n_{i,j}$ stands for $u$ or $d$.
Also the fact that the $\sigma$ is lighter than the $a_0$(980) could then be
easily explained from the smaller hyperfine attraction in the $a_0$(980),
owing to the presence of the (heavier) strange quarks in the latter.

However, notwithstanding the originality and elegance of Jaffe's idea, a
serious problem, which also plagues the many recent yet similar tetraquark
models, is its disregard of unitarisation effects. In other words, the
coupling to physical decay channels, which e.g.\ produce the huge widths of
the $\sigma$ and the $\kappa$, will almost inevitably give rise to real mass
shifts of at least the same order, i.e., of (many) hundreds of MeVs. Therefore,
as long as tetraquarks are not unitarised --- and we are not aware of \em any
\em \/attempt to do so --- such models can only be considered qualitative,
at best.

A completely different description of the light scalars is due to Scadron and
Delbourgo \cite{S82}, based on dynamical chiral-symmetry breaking. In their
Bethe-Salpeter approach, the very same mechanism that makes the pion massless
in the chiral limit, while giving the light quarks their dynamical mass,
straightforwardly leads to a $\sigma$ mass
$m_\sigma=2m_{\mbox{\scriptsize dyn}}$ in the same limit, just like in the
Nambu--Jona-Lasinio \cite{NJL61} model. More recently, Delbourgo and Scadron
\cite{DS95} formulated a similar picture in a non-perturbative and
self-consistent field-theoretic way, via the quark-level linear $\sigma$ model
(QLL$\sigma$M), both for $SU(2)$ and for $SU(3)$, thus predicting a complete
light scalar nonet having masses compatible with present-day experiment. Note
that in this formalism at least some coupled-channel effects are already
accounted for, through meson-loop contributions that recover tree-level
results by construction, via a non-perturbative bootstrap \cite{DS95}.

Another totally alternative view is the generation of at least some of the
light scalars as dynamical resonances of two pseudoscalar (P)  mesons. In
particular, the $a_0$(980) and $f_0$(980) are described as a kind of
$K\bar{K}$ molecules, owing to $K$-$\bar{K}$ potentials that are either
effective or mainly based on $t$-channel vector-meson exchange. In the former
approach, due to Weinstein and Isgur \cite{WI82}, the meson-meson interactions
were extracted from a $q^2\bar{q}^2$ system and couplings to $q\bar{q}$
scalar mesons. In the latter purely mesonic approach, due to Janssen, Pearce,
Holinde and Speth \cite{JPHS95}, even a kind of $\sigma$ was found, though
too light and a bit too broad, with a pole at $(387-i305)$ MeV. However, no
$\kappa$ pole was reported.

A more empirical but also interesting approach is due to Anisovich (V.~V.),
Anisovich (A.~V.), Sarantsev and co-workers \cite{APS96}, who carried out
$K$-matrix analyses of $S$-wave PP scattering data,
thereby identifying the $K$-matrix poles with the bare scalar $q\bar{q}$
states that should follow from quark-antiquark (or gluon-gluon) interactions
only. Although the idea to try to extract information on ``quenched-QCD''
spectra from scattering data is very appealing, and goes way beyond the
traditional, naive approach to meson spectroscopy, the identification of
$K$-matrix poles with bare states puzzles us. Namely, $K$-matrix poles
correspond to the real energies at which the meson-meson phase shift passes
through $90^\circ$, but bare states do not couple at all to the continuum
and so for these there simply is no phase shift. We rather believe the bare
states are at the real energies where (some) $S$-matrix poles end up if one
manages to continuously turn off the coupling to the continuum. We shall
come back to this point below. Anyhow, the distorted scalar nonets inferred
from the analyses in Ref.~\cite{APS96}, missing the $\sigma$ and the $\kappa$,
already hint at problems with the $K$-matrix identification.

A phenomenological chiral model has been developed by Schechter and
collaborators \cite{HSS96}, in which a crossing-symmetric amplitude is
constructed by summing a current-algebra contact term and leading resonance
pole exchanges. Inclusion of ``putative'' light scalar mesons then satisfies
unitarity bounds to well above 1 GeV, and allows good fits to the data.

Another formalism based on effective chiral Lagrangians is the work by
Oller, Oset and Pel\'{a}ez \cite{OO97}, which is often called unitarised
chiral perturbation theory (ChPT). Amplitudes from leading-order or
next-to-leading-order ChPT are unitarised via multichannel
Lippmann-Schwinger equations or other coupled-channel methods. After fitting
the parameters to the available data, a nonet of light scalar mesons shows up
as dynamical resonance poles, though the $f_0$(980) needs a
preexisting bare state in one of the formulations ($N/D$ method).

Recently, a prediction for the $\sigma$ pole was obtained by Caprini,
Colangelo and Leutwyler (CCL) \cite{CCL06}, via Roy equations applied to the
$I\!=\!0$ and $I\!=\!2$ $\pi\pi$ scattering amplitudes. The Roy equations
amount to a twice-subtracted dispersion relation, so that some input from
theory or experiment is needed to fix the subtraction constants, for which
CCL took predictions from ChPT. The resulting $\sigma$ pole position of
$(441^{+16}_{-8}-i272^{+9}_{-12.5})$ MeV is somewhat on the low \cite{PDG08}
side as for its real part. Moreover, the claimed small errors seem too
optimistic, mainly because of the incomplete treatment of the $K\bar{K}$
channel, which opens far below the energy at which the integrals are cut off
(1.4 GeV), but also in view of the uncertainties concerning the true scattering
lengths, the cut-off high-energy ($>1.4$ GeV) tail of the dispersion integral,
and the contributions of higher partial waves.

Already many years earlier, two independent unitarised quark models were
developed and applied to the scalar sector, namely by the Helsinki group, led
by T\"{o}rnqvist \cite{T82}, and several people from Nijmegen \cite{ERMDRR86},
including two of the present authors. In the T\"{o}rnqvist approach, for
each isospin a bare scalar $q\bar{q}$ state in the 1.4--1.6 GeV region is
coupled to all available PP channels, using
$SU(3)$-symmetric coupling constants.  As a result, some dynamical resonances
show up at much lower energies, alongside the unitarised states at the
normal energies for $^{3\!}P_0$ $q\bar{q}$ states. Originally, this only
allowed to generate the $a_0$(980) and $f_0$(980) (Ref.~\cite{T82}, 1st
paper). Much later, inclusion of the Adler-Weinberg zero in the isoscalar case
then also resulted in a light and broad $\sigma$, with pole position
$E=(470-i250)$ MeV (Ref.~\cite{T82}, 2nd and 3rd paper). However, no $\kappa$\
was found in the latter analysis either, which was probably due to the
inclusion of an unphysical Adler zero at a negative value of $s=E^2$
\cite{RKB05}.

In the original Nijmegen approach \cite{ERMDRR86}, one or more confined
$q\bar{q}$ channels were coupled to all available PP as well as vector-vector
(VV) channels, with confinement modelled through a harmonic-oscillator (HO)
potential having flavour-independent spacings. Unitarisation then leads to
a distortion of the bare HO spectrum, and in the scalar case even gives rise
to a doubling of all ground states, resulting in a \em complete \em \/light
scalar nonet, including the $\kappa$, with pole positions 
$(470-i208)$~MeV ($\sigma$), $(727-i263)$~MeV ($\kappa$),
$(994-i20)$~MeV ($f_0$(980)) and $(968-i28)$~MeV ($a_0$(980)). 
These parameter-free predictions from more than 20 years ago are still
close to the present-day world averages. Also note that the same model
predicted earlier \cite{BRRD83} a long controversial $\rho$(1250) resonance,
which was very recently confirmed \cite{SB08} in a multichannel analysis of
$P$-wave $\pi\pi$ data. \\[-8mm]

\section{Resonance-Spectrum-Expansion Model}
\mbox{ } \\[-7mm]
A modern formulation of the coupled-channel model employed in
Ref.~\cite{BRRD83} is the Resonance-Spectrum Expansion (RSE) \cite{BR06a},
in which mesons in non-exotic channels scatter via an infinite set
of intermediate $s$-channel $q\bar{q}$ states, i.e., a kind of Regge
propagators \cite{BR08a}. The confinement spectrum for these bare $q\bar{q}$
states can, in principle, be chosen freely, but in all phenomenological 
applications so far we have used an equidistant HO spectrum, as in
Refs.~\cite{BRRD83} and \cite{ERMDRR86}. Because of the separability of
the effective meson-meson interaction, the RSE model can be solved in closed
form. The relevant Born and one-loop diagrams are depicted in Fig.~1, from
which it is obvious that one can straightforwardly sum up the complete Born
series. For the meson-meson--quark-antiquark vertex functions we take a
delta shell in coordinate space, which amounts to a spherical Bessel function
in momentum space. Such a transition potential represents the breaking of
the string between a quark and an antiquark at a certain distance $r_0$, with
overall coupling strength $\lambda$, in the context of the $^{3\!}P_0$ model.

Spectroscopic applications of the RSE are manifold. In the one-channel
formalism, the $\kappa$ was once again predicted in 2001 (Ref.~\cite{BR01},
1st paper), a year before its experimental confirmation. In the 2nd paper
of Ref.~\cite{BR01}, the low mass of the $D_{s0}^*$(2317) was shown to be
due to its strong coupling to the $S$-wave $DK$ threshold, an
explanation that is now widely accepted. The 3rd paper of Ref.~\cite{BR01}
presented a similar solution to the whole pattern of masses and widths of
the charmed axial-vector mesons.

Multichannel versions of the RSE model have been employed to produce
a detailed fit of $S$-wave PP scattering and a complete light scalar nonet
(Ref.~\cite{BBKR06}, 1st paper), with very few parameters (also see below),
and to predict the $D_{sJ}$(2860) (Ref.~\cite{BBKR06}, 2nd paper), shortly
before its observation was publicly announced.

Finally, the RSE has recently been applied to production processes \cite{BR07}
as well, in the spectator approximation. Most notably, it was shown that the
RSE results in a \em complex \em \/relation between production and scattering
amplitudes (papers 1--3 in Ref.~\cite{BR07}). Successful applications include
the extraction of $\kappa$ and $\sigma$ signals from data on 3-body decay
processes (4th paper in Ref.~\cite{BR07}), the deduction of the string-breaking
radius $r_0$ from production processes at very different energy scales
(5th paper), and even the discovery of signals hinting at new vector charmonium
states in $e^+e^-\to\Lambda_c\bar{\Lambda}_c$ data (6th paper). \\[-6mm]
\begin{figure}[t]
\begin{tabular}{lr}
\resizebox{!}{60pt}{\includegraphics{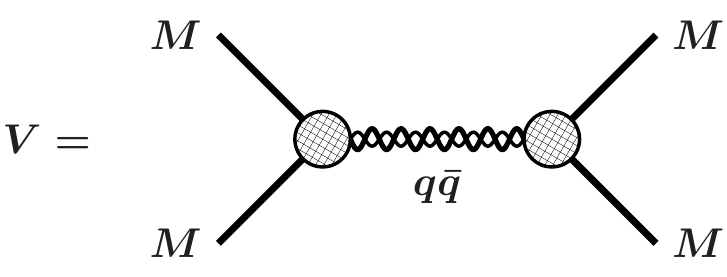}}
&
\hspace*{0.85cm}\resizebox{!}{60pt}{\includegraphics{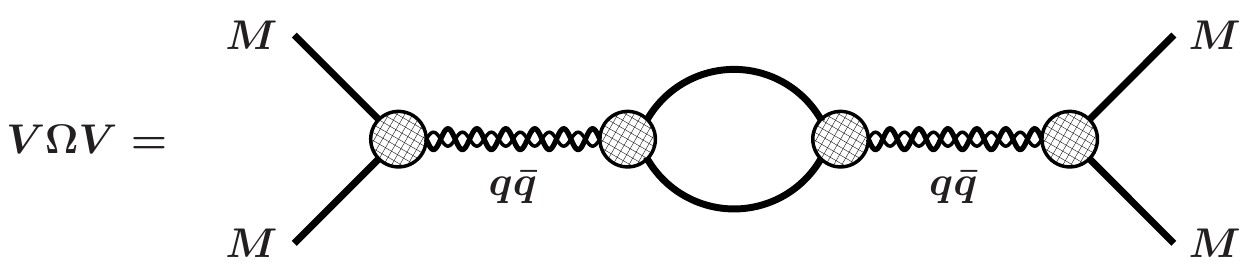}}
\end{tabular}
\caption{\small Born and one-loop term of the RSE effective
meson-meson interaction (see text).}
\end{figure}
\section{Light and Intermediate Scalar Mesons} 
\mbox{ } \\[-18mm]
\subsection{Published results for $S$-wave PP scattering}
In Ref.~\cite{BBKR06}, 1st paper, hereafter referred to as BBKR, two of us
(EvB, GR) together with Bugg and Kleefeld applied the RSE to $S$-wave PP
scattering up to 1.2 GeV, coupling the channels
$\pi\pi$, $K\bar{K}$, $\eta\eta$, $\eta\eta'$, $\eta'\eta'$ for $I\!=\!0$,
$K\pi$, $K\eta$, $K\eta'$ for $I\!=\!1/2$, and $\eta\pi$, $K\bar{K}$,
$\eta'\pi$ for $I\!=\!1$. Moreover, in the isoscalar case both an $n\bar{n}$
and an $s\bar{s}$ channel were included, so as to allow dynamical mixing to
occur via the $K\bar{K}$ channel. The very few parameters, essentially only
the overall coupling $\lambda$ and the transition radius $r_0$, were fitted to
scattering data from various sources, for $I\!=\!0$ and for $I\!=\!1/2$, and to
the $a_0$(980) line shape, determined in a previous analysis, for $I\!=\!=1$.
Moreover, the parameters $\lambda$ and $r_0$ varied less than $\pm10\%$ from
one case to another. Overall, a good description of the data was achieved
(see BBKR for details). Poles for the light scalars were found at
(all in MeV) \\[1.5mm] \indent
$\sigma\!: 530-i226\;, \;\;\;
\kappa\!: 745-i316\;, \;\;\;
f_0(980)\!: 1007-i38\;, \;\;\;
a_0(980)\!: 1021-i47 \;.$ \\[1.5mm]
No pole positions for the intermediate scalar mesons were reported in
BBKR, as the fits were only carried out to 1.1 GeV in the
isovector case, and to 1.2 GeV in the others. Nevertheless, corresponding
poles at higher energies were found, but these were of course quite
unreliable.

In the following, we shall present very preliminary results for fits extended
to higher energies, and with more channels included.

\subsection{Isoscalar scalar resonances with PP and VV channels included}
\parbox[t]{80mm}{
\epsfxsize=80mm
\epsfbox{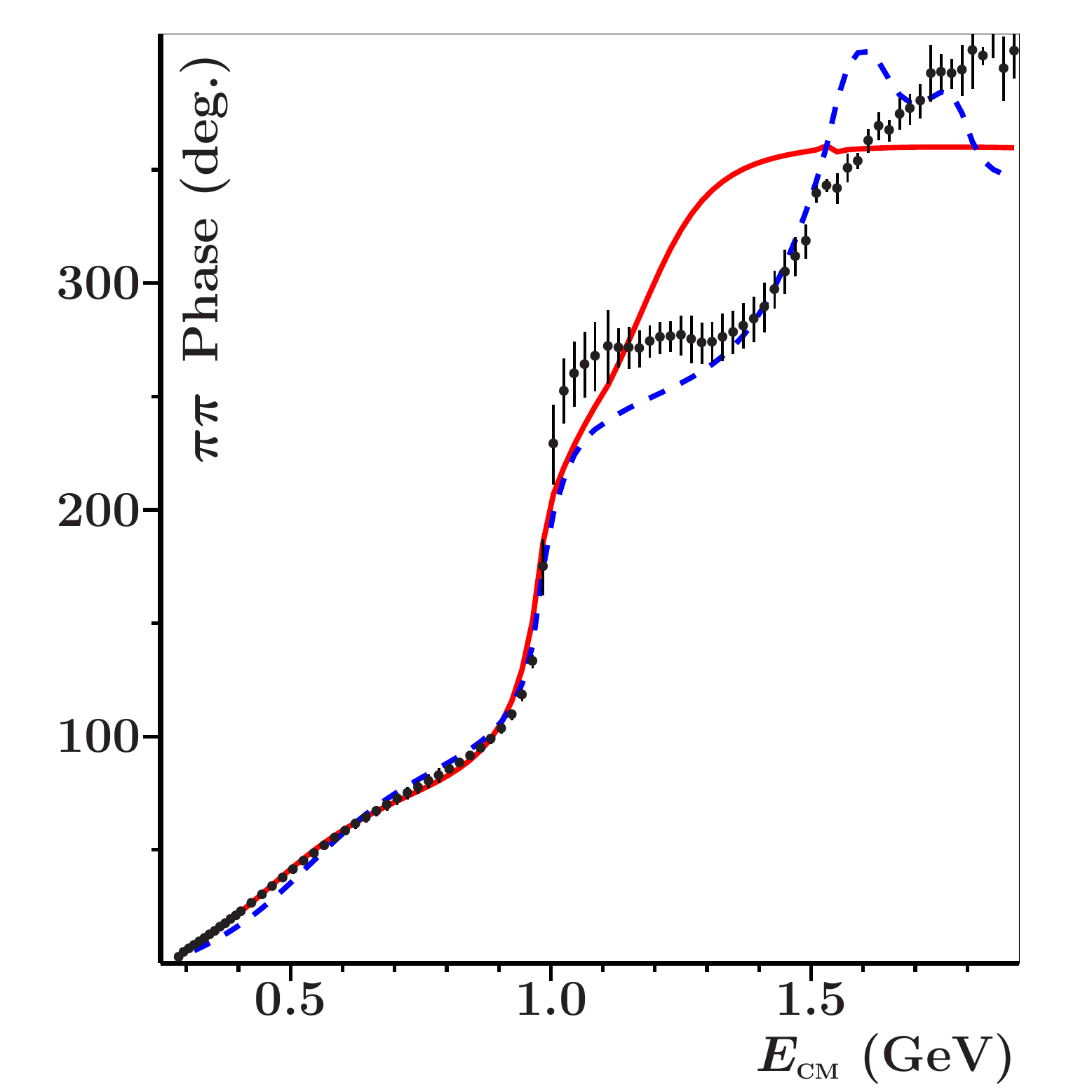} \\ \small\baselineskip 4.5mm
Figure~2: $S$-wave $\pi\pi$ phase shifts: full curve for fit with PP+VV
channels; dashed curve for PP fit from BBKR; data due to Ref.~\cite{BS08}.
} \hfill \parbox[t]{80mm}{\mbox{}\\[-80mm]
For $I\!=\!0$, the VV channels that couple to $n\bar{n}$ and/or $s\bar{s}$
are $\rho\rho$, $\omega\omega$, $K^*\bar{K}^*$ and $\phi\phi$, for both
$L\!=\!0$ and $L\!=\!2$. The corresponding expression for the $T$-matrix is
quite complicated (see Eqs.~(4) and (5) in BBKR). We fit the
parameters $\lambda$ and $r_0$ to sets of $S$-wave $\pi\pi$ phase shifts
compiled by Bugg and Surovtsev \cite{BS08}, which yield a somewhat
larger scattering length than in BBKR, viz.\ $0.21m_\pi^{-1}$.
The result of the fit is is shown in Fig.~2, together with the curve from
BBKR, where only PP channels were included and somewhat
lower data were used just above the $\pi\pi$ threshold. The PP+VV fit is
excellent up to 1 GeV, but clearly lacks structure thereabove. Nevertheless,
inclusion of the VV channels does take care of the unrealistic ``bump''
around the $\eta\eta'$ threshold in the case of the PP fit (up to
1.2~GeV) \hfill of \hfill BBKR \hfill (dashed \hfill curve \hfill in}
\\[0.3mm] Fig.~2). The deficient
behaviour of the PP+VV phase above 1 GeV seems to be mostly due to 
a too low-lying pole for the $f_0(1370)$ and a too broad one for the
$f_0(1500)$.  The first four isoscalar poles we find are (all in MeV) \\[1.5mm]
\indent
$\sigma\!: 456-i234\;, \;\;\;
f_0(980)\!: 994-i57\;, \;\;\;
f_0(1370)\!: 1212-i104\;, \;\;\;
f_0(1500)\!: 1517-i194 \;.$ \\[-8mm]
\subsection{\boldmath{$a_0(980)$} and \boldmath{$a_0(1450)$}}
\mbox{ } \\[-7mm]
In the isotriplet case, we fit $\lambda$ and $r_0$, as well as
the pseudoscalar mixing angle, to the $a_0$(980) line shape, just as in
BBKR, but now with the VV channels 
($\rho K^*$, $\omega K^*$, $\phi K^*$) added. Thus, the quality of the fit
is further improved, and also the fitted mixing angle
$\theta_{PS}=43.7^\circ$ (flavour basis) becomes more realistic.
The poles we find are $(1018-i44)$~MeV (second sheet) for the $a_0$(980)
and $1410-i261$~MeV for the $a_0$(1450), the latter being quite a bit too
broad. \\[-8mm]
\subsection{\boldmath{$K_0^*(800)$} and \boldmath{$K_0^*(1430)$}}
\mbox{ } \\[-7mm]
Including the vector channels ($\rho K^*$, $\omega K^*$, $\phi K^*$) in the
isodoublet sector does not improve the fit, on the contrary. This will
require a detailed investigation, which lies beyond the scope of this
presentation. Thus, we limit ourselves to extend the PP fit from
BBKR up to 1.5~GeV, using the full LASS data, instead of the
ones with the effect of the $K_0^*$(1430) subtracted.
Then, a good fit is obtained, with the very realistic pole positions
$(758-i295)$~MeV for the $\kappa$ and $(1410-i124)$~MeV for the
$K_0^*$(1430). \\[-8mm] 
\section{Conclusions and outlook}
\mbox{ } \\[-7mm]
The preliminary results in this study indicate that a good description of both
the light and the intermediate scalar mesons is feasible in the RSE,
taking into account additional sets of coupled channels that should become
relevant at higher energies. However, for a detailed description of phase
shifts above 1~GeV, it may be necessary to include scalar-scalar channels as
well, e.g.\ to effectively deal with (part of the) $4\pi$ decays in the
isoscalar case. This could also require a procedure to make certain thresholds
less sharp, in order to account for the large widths of some final-state
resonances. 

Support from the {\it Funda\c{c}\~{a}o para a Ci\^{e}n\-cia e a Tecnologia}
\/of the {\it Minist\'{e}rio da Ci\^{e}ncia, Tecnologia e Ensino Superior}
\/of Portugal is acknowledged, cf.\ contract POCI/FP/81913/2007.
One of us (G.R.) thanks the organisers for the kind invitation
to speak at the conference. \\[-7mm]

\end{document}